\documentstyle[preprint,aps]{revtex}
\begin{document}
\draft
\begin{title}
{Theory of SIS tunnelling in cuprates}
\end{title}
\author{A.S. Alexandrov and C. Sricheewin}
\address
{Department of Physics, Loughborough University, Loughborough LE11
3TU, U.K.}

\maketitle
\begin{abstract}
 We show that the  single-particle polaron Green's function 
describes  SIS tunnelling in cuprates, including the absence of Ohm's
law at high voltages, the dip/hump features  in the first derivative
 of the current,   a substantial
 incoherent spectral weight beyond  quasiparticle peaks and   unusual shape  of
 the peaks. 
 The theory allows us to determine the
characteristic phonon frequencies, normal and superconducting gaps,
impurity scattering rate, and the electron-phonon coupling from the
tunnelling data.   

\end{abstract}
\narrowtext

There is a number of    anomalous properties of high-T$_{c}$  cuprates,
which seem to be intrinsic and universal.  The isotope effect on the carrier
mass \cite{guo}, unusually high values
of the static dielectric constants \cite{alebra},  and the  non-Fermi liquid and non-BCS spectral densities observed in the angle-resolved 
photoemission (ARPES) \cite{din,she,sai}, superconductor-
insulator-normal metal (SIN) and superconductor-
insulator-superconductor(SIS) tunnelling  spectra
\cite{pon,brus1,ren,dew,brus2,pmul,kras} are among those. The isotope effect and the
high values  of  the static
dielectric constants  as well as the  optical
spectroscopy \cite{tim} suggest that  electron-phonon interaction is
more than
sufficient to bind  carriers and phonons into small polarons and
bipolarons in highly polarisable  cuprates
\cite{alebra,guo2}.  The bipolaron theory \cite{alemot} provides a natural
explanation of the normal state pseudogap (as a half of the bipolaron
binding energy, $\Delta_{p}$) \cite{ale2,aleray}, NMR linewidth \cite{ale2},  normal state kinetics  
\cite{alemot}, SIN  tunnelling \cite{ale3} and  ARPES \cite{aleden,aleden2}. The theory accounts for two
distinct energy gaps (coherent and incoherent) \cite{aleand} as
observed in the recent tunnelling \cite{brus2,pmul,kras} and Andreev
reflection experiments \cite{yag,brus2}, for  the parameter-free fit of the
critical temperature , upper critical field ,
and specific heat \cite{aleedw}.

In this letter we present the polaron theory of SIS tunnelling in
cuprates.

Within the standard approximation \cite{sup} the 
tunnelling current, $I(V)$,  between two parts of a superconductor separated by an
insulating barrier is proportional to a convolution of the Fourier
component of the single-hole Green's
function (GF),$ G({\bf k}, \omega)$, with itself as
\begin{equation}
I(V)\propto \sum_{\bf k,p} \int_{-\infty}^{\infty} d\omega \Im G({\bf k},
\omega) \Im G({\bf p}, e|V|-\omega),
\end{equation}
where  $V$ is the voltage across the junction. 

A problem of a hole on a lattice  coupled with the bosonic
field of lattice vibrations has a solution in terms of the
coherent (Glauber) states in the strong-coupling limit,
$\lambda > 1 $, where the Migdal-Eliashberg theory \cite{mig,eli}  cannot be applied
due to  the broken translational symmetry \cite{alemaz}.  For
any type of   electron-phonon interaction conserving the on-site
occupation numbers of fermions    the $1/\lambda$ perturbation 
technique yields (at $T=0$) \cite{ale4,alechan}
\begin{equation}
 G({\bf k},\omega)=Z \sum_{l=0}^{\infty}\sum_{{\bf q}_{1},...{\bf q}_{l}}{\prod_{r=1}^{l}|\gamma({\bf
q}_{r})|^{2}\over{(2N)^ll!
(\omega -\sum_{r=1}^{l} \omega({\bf q}_{r})- \epsilon({\bf
  k}+\sum_{r=1}^{l}{\bf q}_{r})+i\delta)}},
\end{equation}
where  $Z=\exp [-(1/2N)\sum_{\bf q}|\gamma({\bf q})|^{2}]$,
$\gamma({\bf q})$ is the matrix element of the interaction with phonons
of the  frequency  $\omega(\bf q)$ \cite{ref} and $\delta->+0$ ($\hbar=c=1$). The
hole energy spectrum, $\epsilon(\bf k)$ is renormalised due to
familiar polaronic narrowing of the band, and (in the superconducting
state) also due to the
interaction with the Bose-Einstein condensation (BEC) of bipolarons
\cite{aleand} as
\begin{equation}
\epsilon({\bf k})= \left[ \xi({\bf k})^2 + \Delta_{c}^2 \right]^{1/2}.
\end{equation}
Here $\xi({\bf k})=Z'E({\bf k})-\mu$ is the renormalised polaron band
dispersion with the chemical potential $\mu$, $E({\bf k})= \sum_{\bf m}t({\bf
m})exp(-i{\bf k \cdot m})$ is the bare (LDA) disperison in a rigid
lattice. The
mass-renormalisation exponent is
\begin{equation}  
Z'={\sum_{\bf m}t({\bf
m})e^{-g^2({\bf m})}\exp(-i{\bf k \cdot m})\over{\sum_{\bf m}t({\bf
m})\exp(-i{\bf k \cdot m})}}
\end{equation}
with 
$g^2({\bf m})=\sum_{\bf q} |\gamma({\bf q})|^{2} [1-\cos ({\bf q \cdot
 a})]$. GF, Eq.(2), is exact in the extreme strong-coupling regime, $\lambda \rightarrow
\infty$ (see, for example, Ref. \cite{sup} page 294).  

Quite different from the BCS superconductor  the chemical
potential $\mu$ is negative in the bipolaronic system, so that the
edge of the single-hole band is found above the
chemical potential  at $-\mu= \Delta_{p}$. Near the edge the parabolic one-dimensional
approximation for the oxygen hole is applied \cite{ale3}, 
compatible with the ARPES data \cite{aleden}
\begin{equation}
\epsilon_{\bf k}\simeq {k_{x}^2\over{2m^*}}+\Delta.
\end{equation}
The 'global' gap $\Delta= (\Delta_p^2+ \Delta_c^2)^{1/2}$ comprises the
incoherent temperature-independent  (normal state) gap $\Delta_p$, and the coherent
(superconducting) gap $\Delta_c(T)$. The coherent gap disappears at $T_c$
together with the condensation \cite{aleand}.

The
continuous-time  Quantum-Monte-Carlo simulations 
     \cite{alekor}  
  show that  
 the $1/\lambda$ perturbation result, Eq.(2), is practically
 exact in a wide range of  the  Fr\"ohlich electron-phonon interaction with
 high-frequency phonons, including the weak-coupling regime. 
Different from the canonical  Migdal-Eliashberg GF there is no
damping  ('defasing')  of low-energy  polaronic excitations in Eq.(2) due to the
electron-phonon coupling alone (because of  the
energy conservation \cite{ale2000}). This coupling
leads to the coherent dressing of low-energy carriers by phonons, which appears 
in GF as phonon sided bands with $l \geq 1$.   On the other hand, the  elastic 
scattering by impurities yields  a finite life-time of the Bloch polaronic
  states. For the sake of   analytical transparency we
  model this scattering as a constant imaginary self-energy, replacing
  $i\delta$ in Eq.(2) by a finite $i\Gamma/2$. In fact, the 'elastic' self-energy has been
  found explicitely  as a  function of  energy and momentum
  \cite{ale3,aleden}. Its   energy/momentum dependence 
  is essential in the subgap region of tunnelling, where it 
  determines the value of the zero-bias conductance. However, it hardly
  plays any role in the peak region and higher voltages, which are of
  our prime interest here.   

Substiting Eq.(2) into the current, Eq.(1), and performing the
intergration with respect to frequency and both momenta (using Eq.(5)), we obtain 
the tunnelling conductance, $\sigma(V)=dI/dV$,
\begin{equation}
\sigma(V) \propto \sum_{l,l'=0}^{\infty} \sum_{{\bf q}_{1},...{\bf
    q}_{l}}\sum_{{\bf q'}_{1},...{\bf q'}_{l'}} {\prod_{r=1}^{l} \prod_{r'=1}^{l'}|\gamma({\bf
q}_{r})|^{2}
|\gamma({\bf q'}_{r'})|^{2}\over{(2N)^{l+l'}l!l'!}}
L\left[e|V|-2\Delta-\sum_{r=1}^{l} \omega({\bf q}_{r})-\sum_{r'=1}^{l'} \omega({\bf q'}_{r'}),\Gamma \right], 
\end{equation}
where $L\left[x, \Gamma\right]=\Gamma/(x^2+\Gamma^2)$. To perform the remaining
integrations and summations, we apply a model analog of the
Eliashberg spectral function $\alpha^2F(\omega)$ by replacing ${\bf q}$-sums,
$\sum_{\bf q}|\gamma({\bf q})|^2 A(\omega({\bf q}))/2N$,
  in Eq.(6) for  the integrals $(g^2/\pi)\int d\omega
L\left[\omega-\omega_{0}, \delta \omega\right]A(\omega)$ for any
arbitrarry function of the phonon frequency $A(\omega({\bf q}))$.  In
this way we introduce the characteristic frequency $\omega_0$ of
phonons strongly coupled with holes, their avarage number $g^2$ in the
polaronic cloud, and their dispersion $\delta \omega$. As long as
$\delta \omega$ is  less than $\omega_0$, we can extend the
integration over phonon frequencies from $-\infty$ to $\infty$ and
obtain 
\begin{equation}
\sigma(V) \propto \sum_{l,l'=0}^{\infty} {g^{2(l+l')}\over{l!l'!}}
L\left[e|V|-2\Delta-(l+l')\omega_0, \Gamma+\delta \omega (l+l')\right].
\end{equation}
By replacing the Lorentzian in Eq.(7) with  the Fourier
integral, we perform the summation over $l$ and $l'$ with the
final result for the conductance as
\begin{equation}
\sigma(V) \propto \int_{0}^{\infty} dt \exp\left[ 2g^2 e^{-\delta
    \omega t}\cos (\omega_0t) -\Gamma t\right] \cos\left[2g^2
    e^{-\delta \omega t} \sin(\omega_{0}t)-(e|V|-2\Delta)t\right].
\end{equation}

From the isotope effect on the carrier mass, phonon densities
of states, experimental values of the  normal state pseudogap, and the residual
resistivity one  estimates the coupling strength $g^2$ to be of
the order of 1  \cite{guo}, the characteristic phonon frequency between 20
and 80 meV, the phonon frequency dispersion about a few tens
meV, the gap $\Delta$ about 30 meV,  and the impurity scattering rate of the order of 10 meV.  
 
SIS conductance, Eq.(8), calculated with the parameters in this
 range is shown in Fig 1,a-d for four different values of the
 coupling. The conductance shape is remarkably different from the case of  BCS
 density of states, both s-wave and d-wave. There is no Ohm's law in
 the normal region, $e|V|>2\Delta$, the dip/hump features (due to 
 phonon sided bands) are  clearly seen already in the first derivative
 of the current,  there is a substantial
 incoherent spectral weight beyond the quasiparticle peak for the
 strong coupling, $g^2 \geq 1$, and there is  unusual shape  of
 the quasiparticle peaks.   All these features as well as the
 temperature dependence of the gap are beyond the BSC theory no matter
 what the symmetry of the gap is. However, they perfectly agree
 with the experimental SIS tunnelling spectra in cuprates
 \cite{brus1,ren,dew,brus2,kras}.
In particular, the theory, Eq.(8) quantitatively describes one of the best
 tunnelling spetra obtained on Bi$_2$Sr$_2$CaCu$_2$O$_{8+\delta}$
 single crystals by the break-junction technique \cite{brus1},
 Fig.2. Some excess zero-bias conductance compared with the
 experiment is due to our approximation of the elastic self-energy.
The exact (energy dependent) self-energy provides an excellent agreement
 in this sub-gap region,  as has been shown in
 Ref. \cite{ale3}. A more recent dynamic conductance of Bi-2212 mesas
 (as shown in Fig.2 of Ref. \cite{kras}) is  almost identical to our
 Fig. 1b as well.
 The unusual shape of the main peaks
 (Fig.1a,b) is a simple consequence of the (quasi) one-dimensional hole
 density of states near the edge of the oxygen band, Eq.(5).  The coherent ($l=l'=0$) contribution to the current with no
 elastic scattering ($\Gamma=0$) is given by
\begin{equation}
I_0 \propto \int_{\Delta}^{\infty}{d
  \epsilon\over{(\epsilon-\Delta)^{1/2}}} \int_{\Delta}^{\infty}{d
  \epsilon'\over{(\epsilon'-\Delta)^{1/2}}}
\delta(\epsilon+\epsilon'- e|V|),
\end{equation}
so that the conductance is a $ \delta$ function
\begin{equation}
\sigma_{0}(V) \propto \delta (e|V|-2 \Delta).
\end{equation} 
Hence, the width of the main peaks in the SIS tunnelling, Fig.1,2
measures directly the elastic scattering rate. 

The disappearance of the quasiparticle sharp peaks
 above T$_c$ in Bi-cuprates has also been explained in the framework of the bipolaron
 theory \cite{ale3,aleden2}. Below T$_c$ bipolaronic Bose-Einstein condensation (BEC)  
provides an effective screening of the long-range
 (Coulomb) potential of impurities, while above T$_c$ the scattering
 rate might increase by many times \cite{aleden2}. This
 sudden increase of $\Gamma$ in the normal state   washes out the
 sharp  peaks
 from the tunnelling and ARPES spectra. 

Finally we would like to comment on a possible role of spin
fluctuations in the tunnelling spectra.  If they play a role, the peak-dip separation  
observed in ARPES and tunnelling should be equal to the resonance peak
energy, $E_{r}$, observed in the spin-flip neutron scattering
\cite{abachub}. However, as
discussed recently in the comprehensive review of the experimental
constraints on the physics of cuprates \cite{guo2},  the peak-dip
separation is nearly independent of T$_c$ or the doping level, while
$E_{r}$  is approximately proportional to T$_{c}$. This controversy as
well as the direct comparison  of the electron-phonon and spin fluctuation interactions \cite{alebra} suggest that
the dip/hump features in ARPES and tunnelling arises from strong
electron-phonon coupling (as originally proposed by one of us
\cite{ale3}). Recently, using high resolution ARPES data in conjuction
with that from neutron, optics and local structural probes, Shen $et$
$al$ \cite{she2} provided direct evidence for strong electron-phonon
coupling being important for pairing.

In conclusion, we have derived the tunnelling SIS conductance for the
strongly coupled electrons and phonons in the (bi)polaronic regime. 
The theory 
describes  SIS tunnelling in cuprates, including 
 the spectral shape in the gap region, and  the dip-hump incoherent  features at 
higher voltages. It allows us to determine  the
characteristic phonon frequencies, the gap,
impurity scattering rate, and the electron-phonon coupling from the
tunnelling data.

The authors greatly appreciate enlightening discussions with Guo-meng
 Zhao and Chris Dent. One of the authors,
 C.Sricheewin, has been supported in this
 work by a grant from the Royal Thai Government.

\centerline{{\bf Figure Captures}}

Fig.1. SIS tunnelling conductance in the bipolaronic superconductor
for different values of the electron-phonon coupling, $g^2$, and
$\Delta=$29 meV, $\omega_0=$ 55 meV, $\delta \omega=$20 meV,  $\Gamma=$ 8.5 meV.

Fig.2. Theoretical conductance of Fig.1b (solid line) compared with the
 tunnelling spetrum obtained on Bi$_2$Sr$_2$CaCu$_2$O$_{8+\delta}$
 single crystals by the break-junction technique \cite{brus1}(dots).


\newpage

\begin{thebibliography}{99}
\bibitem{guo}
G. Zhao, M.B. Hunt,H.  Keller, and K.A. M\"uller, Nature ${\bf 
 385}$, 236 (1997).
\bibitem{alebra}
A.S. Alexandrov and A.M. Bratkovsky, J. Phys.: Condens. Matt. {\bf
  11}, L531 (1999).
\bibitem{din}
H. Ding $et$ $al$, Nature (London) ${\bf 382}$, 51 (1996).
\bibitem{she}
Z.-X. Shen and J.R. Schrieffer, Phys. Rev. Lett. {\bf 78}, 1771
(1997) and references therein.
\bibitem{sai}
N.L. Saini $et$ $al$, Phys. Rev. Lett. {\bf 79}, 3467 (1997).
\bibitem{pon}
Yu.G. Ponomarev $et$ $al$, Solid St. Commun. {\bf 111}, 315 (1999).
\bibitem{brus1}
H. Hancotte $et$ $al$, Phys. Rev. B{\bf 55}, R3410 (1997).
\bibitem{ren}
Ch. Renner $et$ $al$, Phys. Rev. Lett. {\bf 80}, 149 (1998).
\bibitem{dew}
Y. DeWilde $et$ $al$, Phys. Rev. Lett. ${\bf 80}$, 153 (1998).
\bibitem{brus2}
A. Mourachkine, Europhys. Lett. {\bf 49}, 86 (2000).
\bibitem{pmul}
P. M\"uller $et$ $al$, unpublished.
\bibitem{kras}
V.M. Krasnov $et$ $al$, Phys. Rev. Lett. {\bf 84}, 5860 (2000).
\bibitem{tim}
T.\,Timusk {\em et al}, in
{\em Anharmonic Properties of High-$T_{c}$ Cuprates}, 
eds. D.\,Mihailovi\'c {\em et al}, 
(World Scientific, Singapore, 1995), p.171.
\bibitem{guo2}
Gio-meng Zhao, Proceedings of the International Summer School on
Strongly Correlated Systems (Debrecen, Hungary,  September 2000)), to
appear in Phil. Mag (2001).
\bibitem{alemot}
 for reviews see A.S. Alexandrov and N.F. Mott,  {\it Polarons
and Bipolarons} (World Scientific, Singapore, 1995), and 
J.T. Devreese, in
Encyclopedia of Applied Physics, vol. 14, p. 383, VCH Publishers (1996).
\bibitem{ale2}
A.S. Alexandrov,  Physica C {\bf 182},
 327 (1991).
\bibitem{aleray}
A.S. Alexandrov and D.K.  Ray, Phil. Mag. Lett. {\bf 63}, 295, (1991).
\bibitem{ale3}
A.S. Alexandrov, Physica C (Amsterdam) ${\bf 305}$, 46 (1998).
\bibitem{aleden}
A.S. Alexandrov and C.J. Dent, Phys. Rev. B {\bf 60}, 15414 (1999).
\bibitem{aleden2}
A.S. Alexandrov and C.J. Dent, Cond-mat/0012234.
\bibitem{aleand}
A.S. Alexandrov and A.F. Andreev, Cond-mat/0005315.
\bibitem{yag}
Y. Yagil $et$ $al$, Physica C (Amsterdam) {\bf 250}, 59 (1995).
\bibitem{aleedw}
for a recent review see A.S. Alexandrov and P.P. Edwards, Physica C
{\bf 331}, 97 (2000). 
\bibitem{sup}
G.D. Mahan, {\em Many Particle Physics} (Plenum, New York, 1990), p. 793.
\bibitem{mig}
A.B. Migdal, Zh. Eksp. Teor. Fiz. {\bf 34}, 1438 (1958)
(Sov. Phys. JETP {\bf 7}, 996 (1958)).
\bibitem{eli}
G.M. Eliashberg, Zh. Eksp. Teor. Fiz. {\bf 38}, 966 (1960); ibid {\bf
39}, 1437 (1960) (Sov. Phys. JETP {\bf 11}, 696 (1960; {\bf 12}, 1000
(1960)).
\bibitem{alemaz}
A.S. Alexandrov and E.A. Mazur, Zh. Eksp. Teor. Fiz. {\bf 96}, 1773
(1989).
\bibitem{ale4}
A.S. Alexandrov, in {\it Models and Phenomenology for
Conventional and High-temperature Superconductivity} (Course CXXXVI of
the Intenational School of Physics `Enrico Fermi', Varenna, 1997), eds. G. Iadonisi,
J.R. Schrieffer and M.L. Chiofalo (IOS Press, Amsterdam, 1998), p. 309.
\bibitem{alechan}
A.S. Alexandrov and C. Sricheewin, Europhys. Lett., {\bf 51}, 188 (2000).
\bibitem{ref}
In the momentum representation the electron-phonon interaction is
$H_{e-ph}=(2N)^{-1/2} \sum_{\bf q,k}\gamma({\bf q}) \omega({\bf q})
  c_{\bf k}^{\dagger} c_{\bf k-q}(d^{\dagger}_{\bf q}+d_{\bf q})$ with  $c_{\bf k}, d_{\bf q}$ the electron and  phonon
  operators, respectively.
\bibitem{alekor}
A.S. Alexandrov and P.E. Kornilovitch,  Phys. Rev. Lett. ${\bf 82}$,
807 (1999).
\bibitem{ale2000}
A.S. Alexandrov, Phys. Rev. B
 {\bf 61}, 12315 (2000). 
\bibitem{abachub}
A. Abanov, and A.V.Chubukov, Phys. Rev. Lett.
 {\bf 83}, 1652 (1999). 
\bibitem{she2}
Z.X. Shen, A. Lanzara and N. Nagaosa, cond-mat/0102244.








\end{thebibliography}
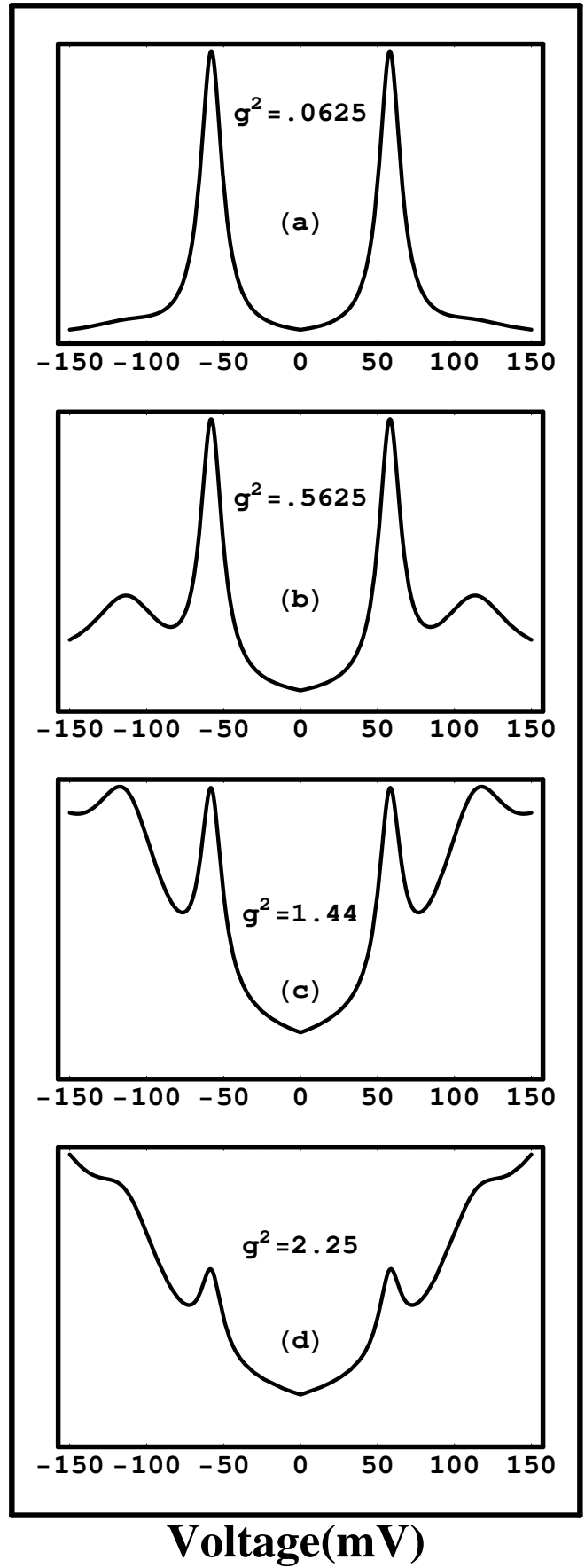
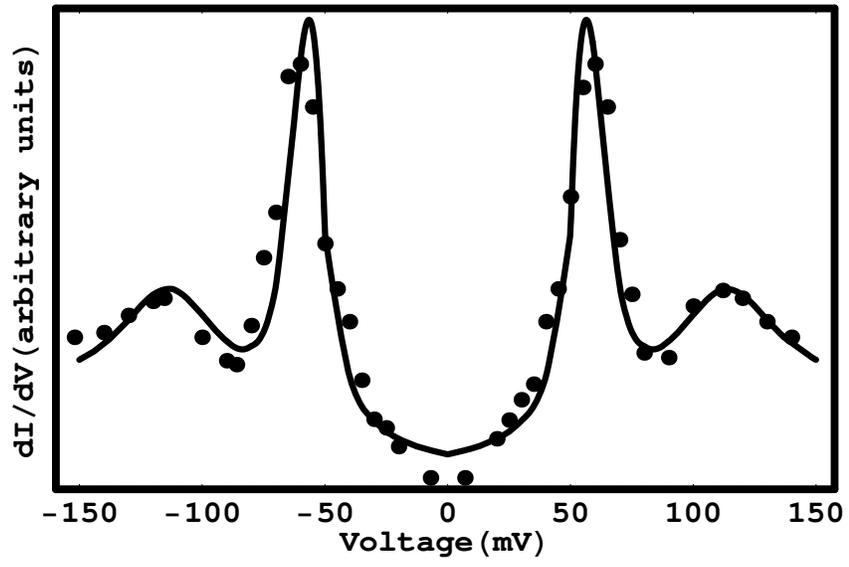
\end{document}